\documentstyle[preprint,aps]{revtex}

\newcommand{\etal}{{\em et al.}}                %       et al.
\newcommand{\beq}{\begin{equation}}
\newcommand{\eeq}{\end{equation}}
\newcommand{\beqar}{\begin{eqnarray}}
\newcommand{\eeqar}{\end{eqnarray}}
\newcommand{\beqars}[1]{\begin{eqnarray*}{#1}}
\newcommand{\eeqars}{\end{eqnarray*}}

\newcommand{\npa}[1]{Nucl.~Phys.~{A#1}}

\def\){\right) }
\def\({\left( }
\def\]{\right] }
\def\[{\left[ }

\begin{document}
%--------------------------------------------------------------
\draft

\title{
Imaging of Sources in Heavy-Ion Reactions
}

\author{
David A.\ Brown\thanks{e-mail:
dbrown@nscl.msu.edu}
and
Pawe\l~Danielewicz\thanks{e-mail:
danielewicz@nscl.msu.edu}
}
\address{
National Superconducting Cyclotron Laboratory and\\
Department of Physics and Astronomy, Michigan State University,
\\
East Lansing, Michigan 48824, USA\\
}
\date{\today}
%\date{November 27, 1996}
\maketitle

\begin{abstract}
Imaging of sources from data within the intensity
interferometry is discussed.  In~the two-pion case,
the~relative pion source function may be determined by
Fourier transforming the correlation function.  In~the
proton-proton case, the~discretized source function may be
fitted to the correlation data.
\end{abstract}


%%%%%%%%%%%%%%%%%%%%%%%%%%%%%%%%%%%%%%%%%%%%%%%%%%%%%%%%%%%%%%%
\narrowtext
Phase interferometry~\cite{arm95} is~capable of delivering 
star images in astronomy, e.g.\ that
of Betelgeuse~\cite{bus90}.  Intensity interferometry, as
is applied in nuclear physics, has primarily been used to
determine radii of stars in astronomy and of particle emitting
regions in heavy-ion reactions.  In heavy-ion reactions, this
determination
has most often been done by fitting the low-momentum
two-particle correlation functions under the assumption of
Gaussian-shaped
emitting regions~\cite{boa90}.  Similarly, source lifetimes 
have been inferred
by considering Gaussian-shaped distributions of emission
times.
However, it~was not always clear that the analyzed data
truly narrows down the lifetime for short-lived sources.
Bertsch~\cite{ber94} and Mr\'{o}wczy\'{n}ski~\cite{mro92} 
noted that an integral over the
low-momentum correlation function can yield the value of the
source function at the zero relative distance.  Beyond this, no
attempts were made to image the source function from the
reaction
data.  Typically, comparison to reaction
simulations is carried out by generating the correlation
functions from simulated events.

In this letter, we investigate the feasibility of direct imaging
of the source from reaction data, within the intensity
interferometry.  Given a~two-particle correlation function,
this represents an~inversion problem.
We consider two examples: that of like-charged-pion and that of
proton-proton interferometry.  In~the like-pion case,
the~relative source function is a~Fourier
transformation of the two-particle correlation function.  
In~the proton case, the~procedure is more involved.
In~this letter, we first discuss the relation of the correlation
function to the source function, then the imaging, and finally
some information contained in the images.

Under the assumption of the approximate independence of
processes leading to two particles in the final state
and the assumption of the weak dependence of the product of
single-particle sources~$\overline{D}$ on
momenta important for correlations, the~two-particle
correlation function may be represented as~\cite{pra90,dan92}
\beq
C_{\bf P}({\bf q}) = {d N_2 / d {\bf p}_1 \, d {\bf p}_2
\over \left( d N_1 / d {\bf p}_1 \right)
\left( d N_1 / d {\bf p}_2 \right) }
 \simeq
\int d{\bf
r} \, |\Phi_{\bf q}^{(-)}({\bf r})|^2 \, S_{\bf P} ({\bf r}).
\label{CPq}
\eeq
Here $S_{\bf P} ({\bf r})$ is the distribution of relative separation 
of emission points for the two particles, in their center of mass.  
In~terms of single-particle sources,
\beq
S_{\bf P} ({\bf r}) = \int d{\bf R} \, dt_1 \, dt_2 \,
\overline{D}(0,{\bf R} + {\bf r}/2, t_1) \,
\overline{D}(0,{\bf R} - {\bf r}/2, t_2) \, .
\label{SPr}
\eeq
The momentum ${\bf P}$ above is the total momentum of the~pair
and ${\bf q}$ is the cm relative momentum.  The~integrations 
in~(\ref{CPq}) and in~(\ref{SPr}) are
over cm variables; the~single-particles sources,
$\overline{D}= D
\left/ \int d{\bf r} \, dt \, D \right. $, are taken in the
pair cm frame.  
The~function~$D$ is the distribution of last
collisions for an~emitted particle in~space, time, and
momentum.  Both $S$~and~$\overline{D}$ are normalized to~1.
For Klein-Gordon fields, with $(\Box + m_\pi^2)
\phi(x) = -j(x)$, $D$~may be written in terms
of single-particle self-energies as
\beq
D({\bf p},{\bf r},t) = {i \over 2 E_p} \, \Pi^< ({\bf p},
E_p, {\bf r}, t)  \exp{\left[ -{1 \over 2 E_p}
\int_t^\infty dt' \, (-2) {\rm Im} \, \Pi^+ \left( {\bf p},
E_p,
{\bf r} + {\bf v}_{\bf p} (t' - t), t' \right) \right]} \, ,
\label{DprK}
\eeq
where $i \Pi^<(x,x') = \langle j(x') \, j(x) \rangle_{\rm
irred}$, and $(-2) {\rm Im} \, \Pi^+ (x,x') = \langle [j(x) ,
j(x')] \rangle_{\rm irred}$.
For the
Schr\"odinger fields, with $\left(i{\partial \over \partial t}
+ {\nabla^2 \over 2 m} \right) \Psi(x) = j(x)$, the~analogous
result is
\beq
D({\bf p},{\bf r},t) = \mp i \Sigma^<({\bf p}, E_p,
{\bf r}, t)  \exp{\left[ - \int_t^\infty dt' \, \Gamma
\left( {\bf p}, E_p, {\bf r} + {\bf v}_{\bf p} (t' - t),
t' \right) \right]} \, ,
\label{DprS}
\eeq
where $\mp i \Sigma^<$ is
the single-particle production rate, $\mp i
\Sigma^< (x,x') = \langle j(x') \, j(x) \rangle_{\rm
irred}$, and $\Gamma$ is the damping rate.
For particles with spin, the modulus of the wavefunction
in~(\ref{CPq}) should be averaged over spin directions.
For low~$r$ within the source, deviations from (\ref{SPr})
could be expected in the presence of short-range repulsion.
Results
(\ref{CPq}), (\ref{DprK}), and (\ref{DprS}) ignore
the final-state refraction.
We discuss the~effects of the Coulomb field of
a~source in a~separate paper~\cite{dan96}.
For~pion-pion correlations, both these effects
and the effects of the pion-pion Coulomb interaction
may be approximately removed directly from the correlation
function~\cite{bay96}.

While the~correlation
function $C$ has an~{\em indirect} dependence on other
quantities,
it is shaped directly by~$S$ in the reaction,
i.e.~by the~relative distribution of emission
points for two particles with similar momenta, in~their
center of mass.
For like particles, $S$ is a symmetric function.  This is
not generally the~case for distinct particles.  Depending
on the circumstances in a~reaction, $S$~may range from
isotropic, for prompt emission, to~strongly elongated
along~${\bf P}$, and possibly even bone-shaped, for emission
from a~long-lived source.  In this last case, the~transverse 
size would be close to the size of a~nucleus at short
distances along ${\bf P}$
and widen at larger distances due to the zigzaging of the source
under the recoil caused by emission.  The~lifetime might be read off from
the elongation of~$S$.

With~(\ref{CPq}), the~goal of the imaging is the determination
of~$S$ given~$C$.  Given that the~interesting part~of~$C$ is its
deviation from~1, we may subtract~1 from both sides
of~(\ref{CPq}) obtaining
\beq
C_{\bf P}({\bf q}) -1 =
\int d{\bf
r} \,
\left(|\Phi_{\bf q}^{(-)}({\bf r})|^2 -1 \right) \,
S_{\bf P} ({\bf r}) =
\int d{\bf
r} \,
K({\bf q}, {\bf r}) \, S_{\bf P} ({\bf r}) ,
\label{K}
\eeq
where $K = |\Phi_{\bf q}^{(-)}|^2-1$.
The~problem of imaging then reduces to the
inversion of~$K$.  A~difficulty may arise from the presence of
a~kernel (or null-space) of~$K$,
i.e.~the~subspace of functions that the operator~$K$ turns to
zero.
The~projection of~$S$ onto the kernel cannot be restored
within imaging.  It~will become apparent that, in the
case of like particles, the kernel of~$K$ is empty.  For unlike
particles, when one of the particles is neutral,
the~imaging may not be able to restore portions of $S$ for
large particle separations as~$K$ approaches
zero for large separations;
a~particular severe situation occurs for $|\Phi_{\bf q}^{(-)}|^2
\simeq 1$, when the whole space of functions becomes the
kernel.

For like-pion pairs, Eq.~(\ref{K}) may be written as
\beq
C_{\bf P} ({\bf q}) - 1 = \int d{\bf r} \cos{\left(2 {\bf q}
\cdot {\bf r} \right) } \, S_{\bf P} ({\bf r}) \, , 
\eeq
if we ignore interactions between the pions.
Given that $S$ is symmetric, the~Fourier transform can be
inverted to yield
\beq
S_{\bf P} ({\bf r}) = {1 \over \pi^3} \int d {\bf q} \,
\cos{\left(2{\bf
q}\cdot{\bf r}\right)} \left( C_{\bf P}({\bf q}) -1 \right) \, .
\label{SPr=}
\eeq
The directions that we use in the analysis of~$S$
are, in the system frame,
outward along the transverse momentum of the
pair, longitudinal along the beam, and the remaining
direction, termed transverse.  With equation~(\ref{SPr=}),
we can find the angular moments of the
source and the correlation function.  If we introduce
$C({\bf q}) = \sqrt{4 \pi} \sum_{\lambda
m} C^{\lambda m} (q) \, {\rm Y}^{\lambda m} (\Omega_{\bf q})$
 and an~analogous representation for~$S$, then
% from~(\ref{SPr=}) 
we get the~relation
\beq
S_{\bf P} ^{\lambda m} (r) = {
(-1)^{\lambda/2} \, 4 \over \pi^2} \int_0^\infty dq \, q^2 \
j_{\lambda} (2 q r)
\, (C_{\bf P}^{\lambda m} (r) - \delta^{\lambda 0} \, \delta^{m 0})
\, .
\label{SPl}
\eeq
Due to the symmetry of~$S$ and~$C$, only even~$\lambda$ appear in
the angular expansion of these functions.  Since both functions
are real, the moments satisfy
$(C^{\lambda m})^* = (-1)^m \,
C^{\lambda \, - m}$.
The~relation~(\ref{SPl}) between the angular
moments may help
in analyzing the three-dimensional data.
In~particular, this~relation shows that
the angle-averaged correlation function $C^{00} (q) \equiv
C(q)$ reflects
the~angle-averaged source $S^{00} (r) \equiv S(r)$,
\beq
r \, S_{\bf P}(r) = {2 \over \pi^2} \, \int_0^{\infty} dq \,
q \, \sin{(2qr)} \, (C_{\bf P} (q) - 1) \, .
\label{rSr}
\eeq

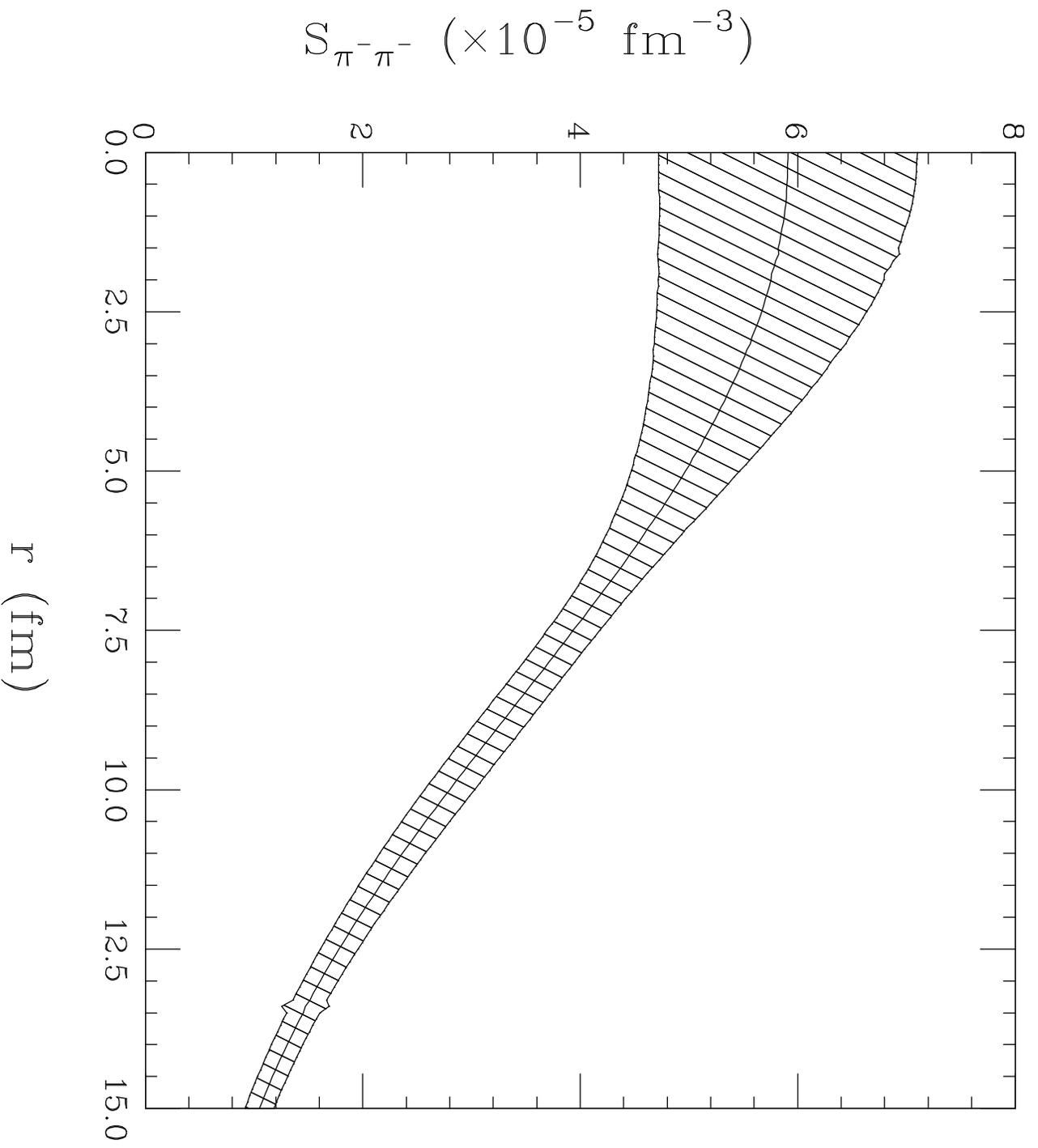
\begin{figure}
\caption{
Relative distribution of emission points for negative pion 
determined from the
correlation data of Ref.~\protect\cite{mis96}.  Prior to the
integration in Eq.~(\protect\ref{rSr}), the~relative momenta in
the correlation function have been increased by~8\% to correct
for the Coulomb effect of the source; similar reduction of the
momenta in the positive-pion correlation function brings the
two correlation functions (or the source functions) into rough
agreement with each other.
The~shaded area represents uncertainty in the source function
associated with the uncertainty in the determination of the
$\pi^-$~correlation function~\protect\cite{mis96} and due to
the choice of $q_{\rm max} \approx 50$~MeV/c.
}
\label{pipi-}
\end{figure}

As~a~specific example of the source extraction,  in
Fig.~\ref{pipi-} we present the relative angle-averaged $\pi^-$
source-function. It is determined by applying~Eq.~(\ref{rSr})  to the
data of Ref.~\cite{mis96} for central 10.8~GeV/c Au~+~Au.  Prior to
the Fourier
transformation in~(\ref{rSr}), the~data were corrected for the
Coulomb interaction between the two pions and between the pions
and
the source~\cite{bay96}.  The~integration in (\ref{rSr})
for~Fig.~\ref{pipi-} was cut off at~$q_{\rm max} \simeq 50$~MeV/c, 
giving a~resolution in
the relative distance in the figure of $\Delta r \sim 1/2q_{\rm
max} \sim 2.0$~fm.  The~largest $r$ that may be considered
is $1/2 \, \Delta q$, where $\Delta q$ is the momentum
resolution of the data ($\Delta q = 5$~MeV/c in~the case
of~\cite{mis96}).

In the general case, the~respective angular moments
of~$C$ and~$S$ are also directly related.  The~spin-averaged
operator,~$K$, only depends on the angle between~${\bf
q}$ and~${\bf r}$, and not on their separate directions.  
Thus,  the~averaged~$K$ may be
expanded: $K({\bf q},{\bf r}) =
\sum_\lambda (2 \lambda +1 )\, K_\lambda (q,r) \, P^\lambda
(\cos{\theta})$. A~relation between the
moments follows from~(\ref{K}),
\beq
C_{\bf P}^{\lambda m}(q) - \delta^{\lambda 0} \, \delta^{m 0}
= 4 \pi \int_0^\infty dr \, r^2 \, K_\lambda (q, r) \,
S_{\bf P}^{\lambda m} (r) \, .
\label{CPl}
\eeq
In the like--nucleon and like--charged--pion cases, the $\lambda = 0$ 
operator is
\beq
K_0 (q, r) = {1 \over 2} \sum_{j s \ell \ell'} (2 j +1) \,
\left( g_{js}^{\ell \ell'} (r) \right)^2 -1 \, ,
\eeq
where $g_{js}^{\ell \ell'}$ is the radial wave function
with outgoing asymptotic
angular momentum~$\ell$.
If~the correlations are of purely Coulomb origin, such as  
between intermediate-mass fragments~\cite{kim91}, 
the~operator is $K_0 (q,r) = \theta(r -r_c) \,
(1 - r_c/r)^{1/2} - 1$ in the classical limit.
Here $r_c$ is the distance of closest approach,
$r_c (q) = 2 \mu \, Z_1 \, Z_2 \, e^2/q^2$.

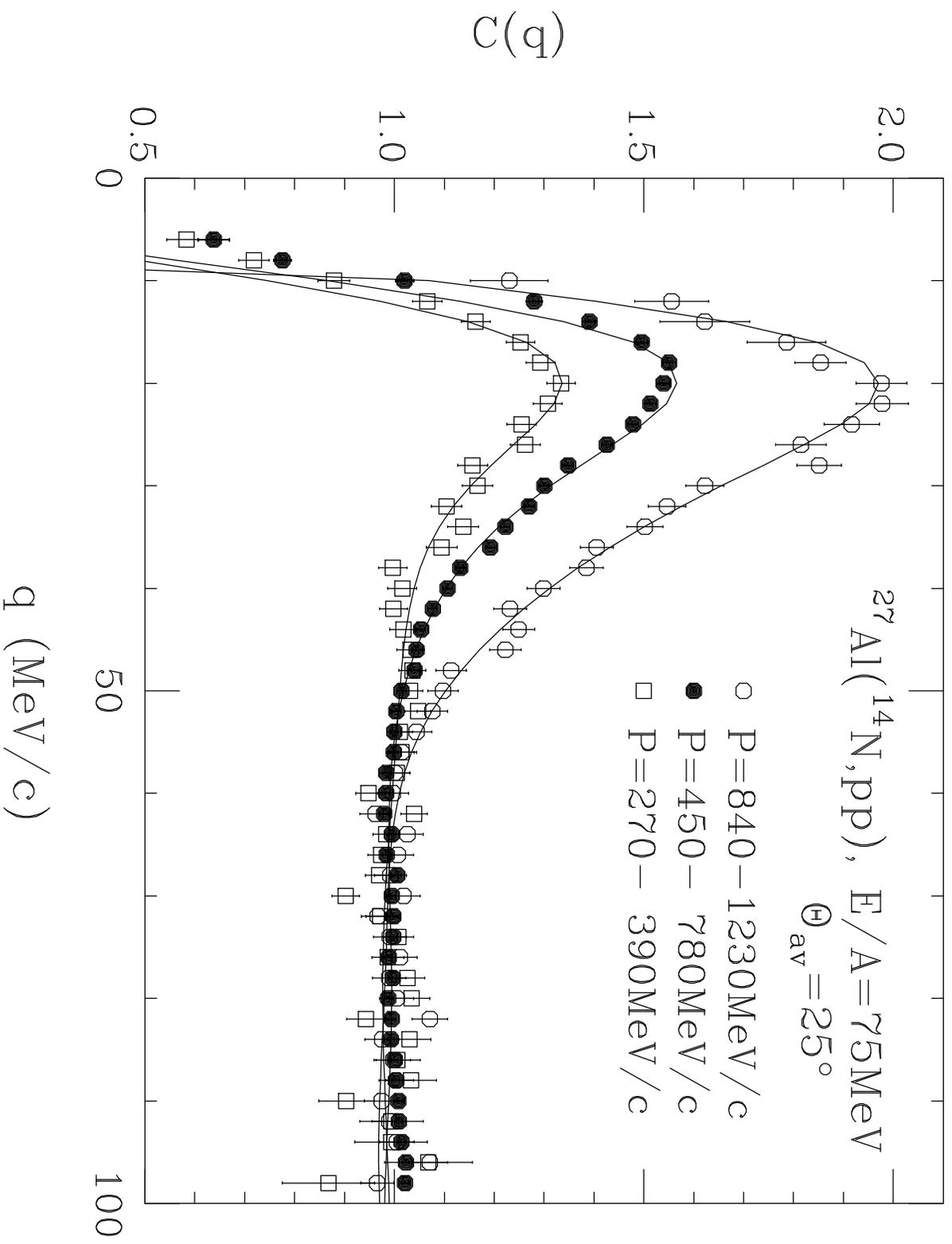
\begin{figure}
\caption{
Two-proton correlation function for the $^{14}$N + $^{27}$Al
reaction at~75~MeV/nucleon.  The~symbols represent
data~\protect\cite{gon91,gon93} for
three gates of total momentum imposed on protons emitted in the
vicinity of $\theta_{\rm lab} = 25^\circ$.  The~lines represent
the correlation function for the extracted sources displayed
in~Fig.~\protect\ref{pps}.
}
\label{pp}
\end{figure}

We determine the source 
%The~source determination may now be carried out 
by discretizing the functions and integrals in
equations such~as
(\ref{CPq}), (\ref{K}), or~(\ref{CPl}), and fitting the
discretized values of~$S$.  We~illustrate this by analyzing the
proton-proton correlation data~\cite{gon93,gon91} from
the 75~MeV/nucleon $^{14}$N + $^{27}$Al reaction,
displayed
in~Fig.~\ref{pp}.

With the data~\cite{gon91} averaged over the ${\bf q}$
directions, we concentrate on the relation~(\ref{CPl}) between
the angle-averaged~$S$ and~$C$.  At~high
relative-momenta, the assumptions leading to~(\ref{CPq}) may
break down.  Further, at~high momentum
correlations within the source, such as associated with the
ansisotropic flow~\cite{kam93}, may play a~role. 
In~addition,
the~event selection for singles in~\cite{gon91} becomes
an~issue.  Thus we~restrict the region
for source determination to less than
$q_{\rm max} \simeq 80$~MeV/c.  Given the issues
of experimental resolution~\cite{gon91}, we~restrict the
region of analysis from~below to~$q > q_{\rm min} = 10$~MeV/c.  
This sets a~lower limit on
the resolution within the source of $1/2(q_{\rm max} - q_{\rm
min}) \simeq 1.4$~fm. Therefore, we~settle with a
determination of the source values
at points separated by a~coarser $\Delta r = 1.8$~fm. 
We~represent the source function as
\beq
S_{\bf P}(r) \simeq \sum_k S_{\bf P} (r_k) \, g (r - r_k) \, ,
\eeq
where $r_k = (k - {1 \over 2}) \Delta r$ and 
$g$ is a~profile function, $g(x) = 1$ for $|x| <
\Delta r /2$, and $g(x) = 0$ otherwise.  Then
from~(\ref{CPl}), we get
\beq
C_{\bf P} (q) - 1 \simeq \sum_k w_k (q) \, S_{\bf P} (r_k) \, ,
\eeq
where $w_k = 4 \pi \int dr \, r^2 \, K_0 (q,r) \, g(r - r_k)$.
We~determine the pp wavefunctions for $\ell, \ell' \leq 2$
from the Schr\"odinger equation with the regularized Reid
soft-core
potential~\cite{sto94}. Finally we minimize $\chi_{\bf P}^2 =
\sum_j ((C_{\bf P}^{\rm exp} (q_j) - C_{\bf P}
(q_j))/\sigma_j)^2$ by varying $S(r_k)$, subject to the
conditions that $S(r_k) \ge 0$ and that~$S$
is normalized to~1.  Initially, we~aimed to
determine
the source up to $r_{\rm max} \sim 30$~fm, but we found that
the fits favored $S$ consistent with zero at higher $r$.
Thus, we~were able to reduce $r_{\rm max}$
to 16.2~fm without an~appreciable worsening in the fits.

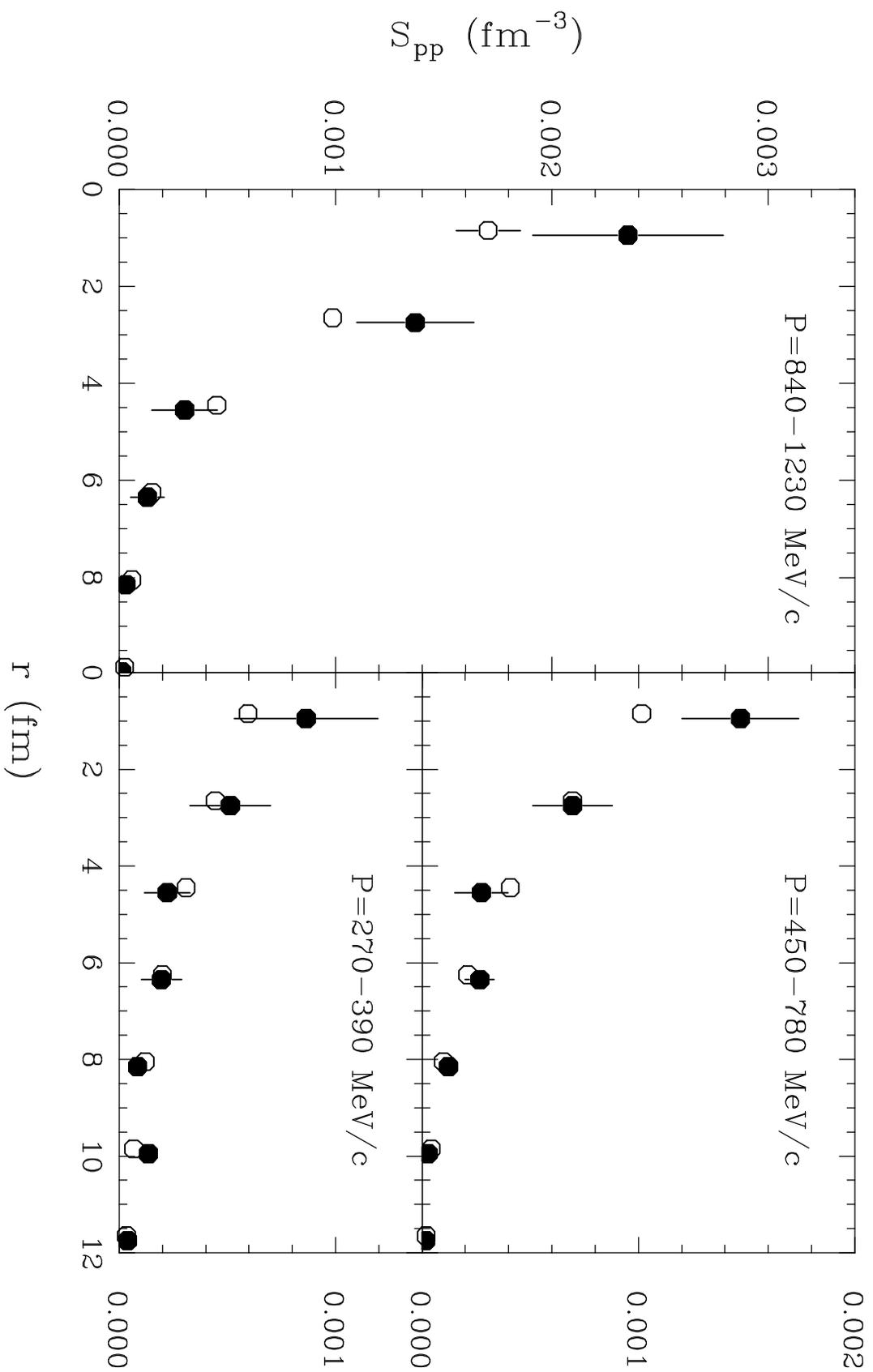
\begin{figure}
\caption{
Relative source function for protons emitted from
the $^{14}$N + $^{27}$Al at~75~MeV/nucleon, in~the vicinity of
$\theta_{\rm lab} = 25^\circ$, within three total momentum
intervals.  Filled circles represent the function extracted from
the data~\protect\cite{gon91,gon93}.  Open circles represent
the function determined within the Boltzmann-equation
model~\protect\cite{dan95}.
}
\label{pps}
\end{figure}

The~source functions
extracted from the data~\cite{gon93,gon91} are shown for the
three total-momentum gates in~Fig.~\ref{pps} together with
the source functions determined directly within
Boltzmann-equation reaction-simulations~\cite{dan95,gon93}.
Unlike what was done in~\cite{gon93}, we initialize the~nuclear matter 
density in the model by solving the Thomas-Fermi equations.
The~errors on the extracted~$S$ include the~uncertainty
from varying $q_{\rm max}$ in the vicinity of~80~MeV/c.
The~values of $\chi^2$ per degree of
freedom are 1.1 for the highest momentum gate, 1.8 for the
lowest, and 4.9 for the intermediate one with the lowest errors
on the data.

We now discuss the relative proton source functions extracted from 
the data and calculated in the model, together with the information 
contained in source functions.  The~change in the relative
distribution of emission points
in~Fig.~\ref{pps}, from a~compact form at high proton
momenta to an~extended form at low momenta, demonstrate the
presence of space-momentum correlations within the
$^{14}$N + $^{27}$Al reaction.
At~intermediate and low momenta,
neither the relative proton distributions from the
data, nor those from the model,
can be well approximated by Gaussians.  This is in contrast
to the distribution in Fig.~\ref{pipi-}, and to a~degree, to
distributions for high momenta in
Fig.~\ref{pps}.
Overall, the~transport model yields distributions in
a~near-quantitative agreement
with the experimental sources.
The~model source seems to systematically underestimate the~values 
from the data only at the shortest~$r$. 

For a~rapid freeze-out, the~single-particle source is given by
$D({\bf p}, {\bf r}, t) \simeq
f({\bf p}, {\bf r}) \, \delta( t - t_0)$ where $f$
is the Wigner function.  Assuming 
weak {\em directional} correlations between the
pair total and relative momenta,
and between the spatial and momentum variables,
the~momentum average of~$S$ approximates
the~relative distribution of emission points for any
two particles
from the reaction, and not just for the particles with close
momenta.  Thus given a~rapid freeze-out, the~relative distribution 
for any two particles is
\beq
{\cal S} ({\bf r}) = {\int d{\bf P} \, d{\bf p} \, d{\bf R} \,
f({\bf P}/2 + {\bf p}, {\bf R} + {\bf r}/2) \,
f({\bf P}/2 - {\bf p}, {\bf R} - {\bf r}/2)
\over
\int d {\bf p}_1 \, d {\bf r}_1 \, f({\bf p}_1 ,{\bf r}_1) \,
\int d{\bf p}_2 \, d {\bf r}_2 \, f({\bf p}_2 ,{\bf r}_2) }
\, .
\label{sr}
\eeq
We now rewrite and expand the expression in the numerator
in~(\ref{sr}):
\beqar
\nonumber
\lefteqn{
f({\bf P}/2 + {\bf p}, {\bf R} + {\bf r}/2) \,
f({\bf P}/2 - {\bf p}, {\bf R} - {\bf r}/2)
} \hspace{2em} \\[.5ex] \nonumber
& = & \int d{\bf r}_1' \, f({\bf p}_1 ,{\bf r}_1') \,
 \int d{\bf r}_2' \, f({\bf p}_2 ,{\bf r}_2') \,
{
f({\bf P}/2 + {\bf p}, {\bf R} + {\bf r}/2) \,
f({\bf P}/2 - {\bf p}, {\bf R} - {\bf r}/2)
\over
\int d{\bf r}_1' \, f({\bf P}/2 + {\bf p} ,{\bf r}_1') \,
 \int d{\bf r}_2' \, f({\bf P}/2 - {\bf p} ,{\bf r}_2')
} \\[1ex] \nonumber
& = & \int d{\bf r}_1' \, f({\bf p}_1 ,{\bf r}_1') \,
 \int d{\bf r}_2' \, f({\bf p}_2 ,{\bf r}_2') \,
\left(1 + {\bf p} \, {\partial \over \partial {\bf p}'}
+ \ldots \right) \\[1ex]
& & \hspace{1em} \times \left.
{
f({\bf P}/2 + {\bf p}', {\bf R} + {\bf r}/2) \,
f({\bf P}/2 - {\bf p}', {\bf R} - {\bf r}/2)
\over
\int d{\bf r}_1' \, f({\bf P}/2 + {\bf p}' ,{\bf r}_1') \,
 \int d{\bf r}_2' \, f({\bf P}/2 - {\bf p}' ,{\bf r}_2')
} \,
\right|_{{\bf p}' = 0} \, .
\label{ff}
\eeqar
The gradient term must be proportional to a~combination of the
vectors ${\bf P}$, ${\bf r}_1$, and ${\bf r}_2$.  For 
weak directional correlations, it then averages to zero under
the integration in~(\ref{sr}).  Inserting~(\ref{ff})
into~(\ref{sr}) and keeping the leading term, we~obtain
\beq
{ \cal S} ({\bf r}) \simeq {1 \over N^2} \int d{\bf p}_1 \,
d{\bf p}_2 \, {d N \over d {\bf p}_1} \,
\, {d N \over d {\bf p}_2} \, \gamma_{P} \,
S_{\bf P}({\bf r} + {\bf n}_{\bf P} \, (\gamma_{P} - 1) ({\bf
n}_{\bf P} \, {\bf r})) \, ,
\label{rs}
\eeq
where $N$ is particle multiplicity and ${\bf n}_{\bf P} ={\bf
P}/P$. The~argument of~$S_{\bf P}$ has been written in
the cm frame of an emitted~pair and $\gamma_{P}$ is the Lorentz
factor for the transformation from the system frame to the pair
cm frame.

Generally the~relative distribution of emission
points for any two particles at~$r \rightarrow 0$
gives an~average freeze-out density when multiplied by~$N-1$.
If~the assumptions
above are valid, this density may be obtained by multiplying
the~average~(\ref{rs}) of~$S_{\bf P}(r \rightarrow 0)$ by~$N-1$.
The~transport
calculations~\cite{dan95} indicate that the measured~\cite{gon91}
coincidence cross sections for the $^{14}$N + $^{27}$Al
reaction are dominated by rather central
collisions with~$b \sim 2.8$~fm.  
The chance of detecting two particles at wide angles simultaneously is 
large only for these collisions.
The~rms nucleon cm momentum in central collisions is $\sim
185$~MeV/c and  at~25$^\circ$ this corresponds to~a~lab momentum~of 
$\sim 320$~MeV/c for the nucleon or $\sim 640$~MeV/c
total for the~pair.  Thus, the~results for the
intermediate-momentum gate in~Fig.~\ref{pps} best
represent the~average situation in the~central collision.
Presuming that the relative  spatial distributions of other 
particles to protons is similiar to that of two protons,
we arrive at an~average
nuclear density in the
vicinity of any emitted proton of $17 \times S(r \rightarrow 0)
\simeq 17 \times 0.0015$~fm$^{-3} = 0.025$~fm$^{-3} = 0.16\,
n_0$.  Here we assume the participants to have a total mass of
18 following from the fireball geometry at~$b \approx
2.8$~fm. The~directional space-momentum correlations
anticipated in collisions due to a~collective motion, to
shadowing, or to~emission that is most likely not instantaneous,
make this value an~upper limit on the freeze-out
density.

Irrespective of any correlations or of the validity of
instantaneous freeze-out, the~product of the $r \rightarrow 0$
source-function and the momentum distribution yields the space
average of the phase-space occupancy at freeze-out,
\beq
\langle f ({\bf p}) \rangle = {(2 \pi)^3 \over 2 s+1}
\, {E_p \over
m} \, {d N \over d{\bf p}} \, S_{2 {\bf p}} (r \rightarrow 0)
\, ,
\label{f}
\eeq
(see also~\cite{ber94}).
Refs.~\cite{gon91,gon93} only give the inclusive proton
cross-sections in the $^{14}$N + $^{27}$Al reaction
at two angles.  Furthermore, 
the cross sections include large contributions from
peripheral events.
Under these circumstances, we~instead use the~thermal
distribution $dN_{\rm th}/d{\bf p}
\propto 1/ (z^{-1} \, {\rm e}^{p^2/2mT} + 1)$
for the central events in~formula~(\ref{f}).
Here, $z$~is
set from the requirement of maximum entropy.  For $\sim 9$
participant protons at~$b = 2.8$~fm in the $^{14}$N + $^{27}$Al
reaction, this~requirement
gives $z \sim 1.10$ and $T \approx 10.2$~MeV.
Equation~(\ref{f}) can now be used to determine the phase-space
average of the occupancy
at freeze-out, $\langle f \rangle = \int d{\bf p} \, (\langle f
({\bf p})\rangle )^2 \, /
\int d{\bf p} \, \langle f ({\bf p})\rangle $, and to estimate
the entropy per nucleon,
\beqars
{S \over A} \approx - \, {\int d{\bf p} \left(\langle f 
({\bf p}) \rangle\, \log{\left(\langle f ({\bf p})\rangle\right)} -
\left(1 - \langle f ({\bf p})\rangle \right) \,
\log{\left(1 - \langle f ({\bf p})\rangle\right)} \right)
\over \int d{\bf p} \, \langle f ({\bf p})\rangle} \, .
\eeqars
Use of the thermal momentum distribution 
yields $\langle f \rangle \approx 0.23$ and $S/A
\approx 2.7$ for the $^{14}$N + $^{27}$Al reaction.  
For a~distribution with nonequilibrium
features, these values should represent the~lower limit on the
average occupation
and the upper limit on the entropy.  Indeed, for the source 
from the transport model and a~thermal 
distribution, we find
an~entropy about 0.5 per nucleon higher than
the entropy calculated directly within the model.

We have shown how to determine the relative source functions
for particles directly
from correlation data.  When going beyond the Gaussian
fitting, direct Fourier inversion, or 
simply fitting the source,
%the plain discretization of the source, 
the~singular value decomposition~\cite{pre86} might be useful.
This~method allows one to determine what kind of source
parametrization can be narrowed  practically down by the
correlation data.
%  We~hope that, with our investigation, 
%we have extended the amount of information
%accessible within the heavy-ion reactions.

\acknowledgements
The authors thank Dariusz Mi\'skowiec and Peter Braun-Munzinger
for providing them the pion correlation data in a~numerical form.
They further acknowledge the
conversations with colleagues,
that have planted the seeds for this work, in~particular with
Herbert Str\"obele, George Bertsch, and Scott Pratt.
Finally, the~authors thank Volker Koch for reading the
manuscript.
This
work was partially supported by the National Science Foundation
under Grant PHY-9403666 and
by the Department of
Energy under Grant FG06-90ER40561.

\newpage

%--------------------------------------------------------------
\end{document}